\documentclass[
    ,final            % use final for the camera ready runs
  ]
  {aipproc}

\layoutstyle{6x9}
\usepackage{amssymb}

\begin{document}

\title[Reconstructing the interaction term]{Reconstructing the interaction term
between dark matter and dark energy}


\author{Freddy Cueva and Ulises Nucamendi}
{address={Instituto de F\'{\i}sica y Matem\'aticas\\
Universidad Michoacana de San Nicol\'as de Hidalgo \\
Edificio C-3, Ciudad Universitaria, CP. 58040\\
Morelia, Michoac\'an, M\'exico} }

\begin{abstract}
We apply a parametric reconstruction method to a homogeneous,
isotropic and spatially flat Friedmann-Robertson-Walker (FRW)
cosmological model filled of a fluid of dark energy (DE) with
constant equation of state parameter interacting with dark matter
(DM). The reconstruction method is based on expansions of the
general interaction term and the relevant cosmological variables in
terms of Chebyshev polynomials which form a complete set orthonormal
functions. This interaction term describes an exchange of energy
flow between the DE and DM within dark sector. To show how the
method works we do the reconstruction of the interaction function
expanding it in terms of only the first three Chebyshev polynomials
and obtain the best estimation for the coefficients of the expansion
as well as for the DE equation of the state constant parameter $w$
using the type Ia Supernova SCP Union data set (307 SNe-Ia). The
preliminary reconstruction shows that in the best scenario there is
an energy transfer from DM to DE which worsen the problem of the
cosmic coincidence in comparison with the $\Lambda$CDM model. We
conclude that this fact is an indication of a serious drawback for
the existence of such interaction between dark components.
\end{abstract}

\maketitle

%%%%%%%%%%%%%%%%%%%%%%%%%%%%%%%%%%%%%%%%%%%%
%% MAINMATTER
%%%%%%%%%%%%%%%%%%%%%%%%%%%%%%%%%%%%%%%%%%%%

%\subsection{Introduction}
In the last years the accelerated expansion of the universe has now
been confirmed by several independent observations including those
of high redshift $(z\leq 1)$ type Ia Supernovae (SNeIa) data at
cosmological distances \cite{Perlmutter1999}-\cite{Kowalski2008}.
This has been verified by precise measurements of the power spectrum
of the cosmic microwave background (CMB) anisotropies
\cite{Balbi2000}-\cite{Spergel2007} and the galaxy power spectrum
\cite{Tegmark2004}. To explain these observations it has been
postulated the existence of a new and enigmatic component of the
universe so-called dark energy (DE). Recent observations
\cite{Hinshaw2009}-\cite{Rozo2009} show that if it is assumed a dark
energy (DE) equation of state (EOS) with constant parameter $w =
P_{DE}/ \rho_{DE}$, then there remains little room for departure of
DE from the cosmological constant, since $\mid 1+w \mid <0.06$ at
the $1\sigma$ confidence level (C.L.). In addition these
observations indicate that our universe is flat and it consists of
approximately $70\%$ of Dark Energy (DE), $25\%$ of Dark Matter and
$5\%$ of barionic matter.

However the cosmological constant model has two serious problems:
the first of them is the \textit{cosmological constant problem}
\cite{Weinberg1989}-\cite{Padmanabhan2003} which consists in why the
observed value of the Cosmological Constant $\rho_{\Lambda}^{obs}
\sim (10^{-12} \,\,\textrm{Gev})^{4}$ is so-small compared with the
theoretical value $\rho_{\Lambda}^{Pl} \sim (10^{18}
\,\,\textrm{Gev})^{4}$ predicted from local quantum field theory if
we are confident in its application to the Planck scale?. The second
problem is the so named \textit{The Cosmic Coincidence problem}
\cite{Steinhardt1997}-\cite{Zlatev1999} consisting in why, in the
present, the energy density of DE is comparable with the density of
dark matter (DM) while the first one is subdominant during almost
all the past evolution of the universe?.

Recently, in order to solve the \textit{The Cosmic Coincidence
problem}, several researchers have considered a possible
phenomenological interaction between the DE and DM components
\cite{Campo2008}-\cite{Amendola2007}. Some of these studies have
claimed that, for reasonable and suitably chosen interaction terms,
the coincidence problem can be significantly ameliorated in the
sense that the rate of densities $r \equiv \rho_{DM}/ \rho_{DE}$
either tends to a constant or varies more slowly than the scale
factor, $a(t)$, in late times.

While it is not totally clear if an interaction term can solved the
\textit{The Cosmic Coincidence problem}, we can yet put constraints
on the size of such assumed general interaction using recent
cosmological data. We do this postulating the existence of an
general nongravitational interaction between the two dark
components. We introduce phenomenologically this interaction term
$Q$ into the equations of motion of DE and DM, which describes an
energy exchange between these components
\cite{Campo2008}-\cite{Amendola2007}. In order to reconstruct the
interaction term $Q$ as a function of the redshift we expand it in
terms of Chebyshev Polinomials which constitute a complete
orthonormal basis on the finite interval [-1, 1] and have the nice
property to be the minimax approximating polynomial (this technique
has been applied to the reconstruction of the DE potential in
\cite{Simon2005}-\cite{Martinez2008}). At the end we do the
reconstruction using the observations of type Ia Supernova SCP Union
data set (307 SNe-Ia) \cite{Kowalski2008}. In our reconstruction
process we assume a DE equation of state parameter $w = constant$.

The organization of this paper is a follow. In the second section we
introduce the equations of motion of the DE model interacting with
DM and the reconstruction scheme of the interaction term. In the
third section we briefly describe the application of the type Ia
Supernova data cosmological test and the priors used in this one. In
the last section we present the results of our reconstruction and
the best estimated values of the parameters fitting the
observations. Finally, we discuss our main results and present our
conclusions.

    \paragraph{\textbf{Equations of motion for dark energy interacting with dark matter}}

The background metric is described by the flat
Friedmann-Robertson-Walker (FRW) metric as supported by the
anisotropies of the cosmic microwave background (CMB) radiation
measured by the WMAP experiment
\begin{equation}
\label{eq:Omega1} ds^{2}=-dt^{2} + a^{2}(t) \left( dr^{2} +
r^{2}d\Omega^2 \right ),
\end{equation}
where $a(t)$ is the scale factor and $t$ is the cosmic time.

We assume an universe formed by four components: the barionic matter
density $\rho_{b}(z)$, the radiation density $\rho_{r}(z)$, the DM
density $\rho_{DM}(z)$ and the DE density $\rho_{DE}(z)$, where the
variable $z$ represents the redshift. The DE equation of state is
assumed as $P_{DE} = w \rho_{DE}$ where $w$ is a constant where as
for the dark matter we have $P_{DM} = 0$.

Moreover all these constituents are interacting gravitationally and
additionally only the components $\rho_{DE}$ and $\rho_{DM}$
interact nongravitationally through an energy exchange between them
mediated by the interaction term $Q(z)$. As we know the solutions
for the barionic matter and radiation density are respectively:
\begin{equation}
\label{eq:Omega10}
\hat{\Omega}_{b}(z) \equiv \frac{\rho_{b}}{\rho_{crit}^{0}} =
\Omega_{b}^{0}{(1+z)}^{3},\\
\end{equation}
\begin{equation}
\label{eq:Omega11}
\hat{\Omega}_{r}(z) \equiv \frac{\rho_{r}}{\rho_{crit}^{0}}=
\Omega_{r}^{0}{(1+z)}^{4},\\
\end{equation}
where $H_0$ is the Hubble constant, $\rho_{crit}^{0} \equiv
3H^2_0/8\pi G$ is the critical density today and $\Omega_{b}^{0}
\equiv \rho_{b}^{0}/\rho_{crit}^{0}$, $\Omega_{r}^{0} \equiv
\rho_{r}^{0}/\rho_{crit}^{0}$ are respectively the density
parameters of barionic matter and radiation at the present. The
energy conservation equations for both dark components are:
\begin{eqnarray}
\label{eq:Omega4} \frac{\mathrm{d{\rho}_{DM}}}{\mathrm{d}z} -
\frac{3}{1+z} \rho_{DM}=
\frac{-Q(z)}{(1 + z) \cdot H(z)},\\
\label{eq:Omega5} \frac{\mathrm{d{\rho}_{DE}}}{\mathrm{d}z} -
\frac{3(1+w)}{1 + z} \rho_{DE} = \frac{Q(z)}{(1+z) \cdot H(z)},
\end{eqnarray}
here $H(z)$ is the Hubble parameter. We complete the equations of
motion with the first Friedmann equation,
\begin{equation}
\label{eq:hubble1} H^{2}\left(z \right) = \frac{8\pi G}{3}
\left(\rho_{b} + \rho_{r} + \rho_{DM} + \rho_{DE} \right).
\end{equation}
Using (\ref{eq:Omega10}) and (\ref{eq:Omega11}) we write
(\ref{eq:hubble1}) as
\begin{equation}
\label{eq:hubble2} H^{2}\left(z \right) = H^2_{0}
\left[\Omega_{b}^{0}{(1+z)}^{3} + \Omega_{r}^{0}{(1+z)}^{4} +
\hat{\Omega}_{DM}(z) + \hat{\Omega}_{DE}(z) \right],
\end{equation}
where we use the definitions $\hat{\Omega}_{DM}(z)\equiv
\rho_{DM}/\rho^{0}_{crit}$, $\hat{\Omega}_{DE}(z)\equiv
\rho_{DE}/\rho^{0}_{crit}$. We do the parametrization of the $Q(z)$
coupling in terms of the Chebyshev polynomials, which form a
complete set of orthonormal functions on the interval $[-1, 1]$.
They also have the property to be the minimax approximating
polynomial, which means that has the smallest maximum deviation from
the true function at any given order. We can then expand $Q$ in the
redshift representation as:
\begin{equation}
\label{eq:Omega8} Q(z) = \sum_{n=0}^{N}\lambda_{n} T_{n}(z)\cdot
H(z)\cdot (1+z)^{3},
\end{equation}
where $T_{n}(z)$ denotes the Chebyshev polynomials of order $n$ with
$n \in [0,N]$ and $N$ a positive integer. $H(z)$ represents the
Hubble parameter and $\lambda_{n}$ are real free parameters
representing the coefficients of the linear expansion. We introduce
(\ref{eq:Omega8}) in (\ref{eq:Omega4}) and (\ref{eq:Omega5}) and
integrate both equations obtaining,
\begin{eqnarray}
\label{eq:Omega12} \hat{\Omega}_{DM}(z) &=& (1+z)^{3}\left[
{\Omega_{DM}^0} - \frac{z_{max}}{2} \sum_{n=0}^{N} \hat{\lambda}_{n}
\int_{-1}^{x}
\frac{T_{n}(\tilde{x})}{(a+b\tilde{x})} d\tilde{x} \right],\\
\label{eq:Omega13} \hat{\Omega}_{DE}(z) &=& (1+z)^{3(1+w)}\left[
{\Omega_{DE}^0} + \frac{z_{max}}{2} \sum_{n=0}^{N} \hat{\lambda}_{n}
\int_{-1}^{x} \frac{T_{n}(\tilde{x})}{(a+b\tilde{x})^{(3w+1)}}
d\tilde{x} \right]
\end{eqnarray}
where we have defined the dimensionless coefficients
$\hat{\lambda}_{n} \equiv \lambda_{n}/\rho^{0}_{crit}$ and the
quantities $a = 1\,+\, z_{max}/2$, $ b = z_{max}/2$, $x = 2z/z_{max}
- 1$ where $z_{max}$ is the maximum redshift at which observations
are available so that $x \in [-1, 1]$ and $ \vert T_{n}(x)
\vert\leq 1$, \,for all $n \in [0,N]$.\\
\\
As we can see from (\ref{eq:Omega12}) and (\ref{eq:Omega13}), the
Hubble parameter (\ref{eq:hubble2}) depends of the parameters
($H_0$, $\Omega_{b}^0$, $\Omega_{r}^0$, $\Omega_{DM}^0$,
$\Omega_{DE}^0$, $w$) and the coefficients $\hat{\lambda}_{n}$. In
general we must take $N\rightarrow \infty$ but in the practice we
cut to some finite integer. To simplify our analysis and to show how
the method works we do the reconstruction taking the first three
polynomials $n = 0, 1, 2$ \,($N=2$), which are:
\begin{eqnarray}
\label{eq:Omega14} T_{0}(x) = 1 \,, \quad\quad T_{1}(x) = x \,,
\quad\quad T_{2}(x) = 2x^{2} - 1  \,,
\end{eqnarray}
using these polynomials we find closed solutions for the expressions
(\ref{eq:Omega12}) and (\ref{eq:Omega13}), and finally from
(\ref{eq:hubble2}) we can write the dimensionless Hubble parameter
$\tilde{H}^2 \equiv H^2/H^2_0$ as
\begin{equation}
\label{hubblefinal} \tilde{H}^{2}(z) = \Omega_{b}^{0}{(1+z)}^{3} +
\Omega_{r}^{0}{(1+z)}^{4} + \hat{\Omega}_{D}(z) \,,
\end{equation}
where $\hat{\Omega}_{D}$ denotes the sum of the density parameters
of both dark components,
\begin{equation}
\hat{\Omega}_{D} (z) \equiv \hat{\Omega}_{DM}(z, \Omega_{DM}^0,
\hat{\lambda}_{0}, \hat{\lambda}_{1}, \hat{\lambda}_{2}) +
\hat{\Omega}_{DE}(z, \Omega_{DE}^0, w, \hat{\lambda}_{0},
\hat{\lambda}_{1}, \hat{\lambda}_{2})  \,.
\end{equation}

\begin{center}
\begin{figure}
  \includegraphics[width=9cm]{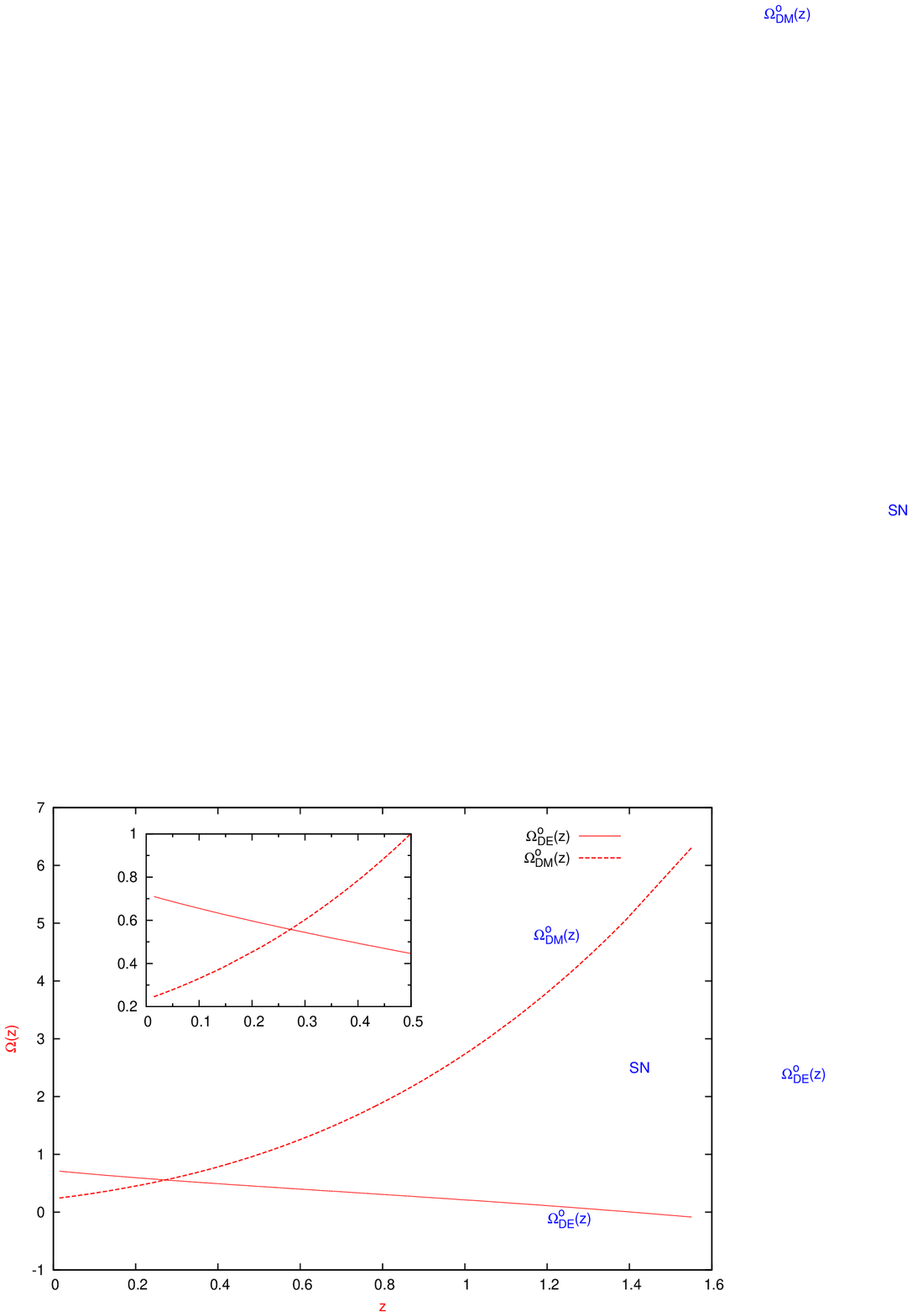}
  \caption{Reconstruction for the density parameters
  $\Omega_{DM}(z), \, \Omega_{DE}(z)$ as a function of the redshift
  for a spatially flat universe with dark energy interacting with dark matter.
  The reconstruction is derived from the best estimation obtained using the type
  Ia Supernova SCP Union data set sample. $z_{max} = 1.551$ corresponding to the
  farthest Supernova in the sample used. Note that the coincidence cosmic problem persists}
  \label{Mixed_SN}
\end{figure}
\end{center}

\begin{center}
\begin{figure}
  \includegraphics[width=9cm]{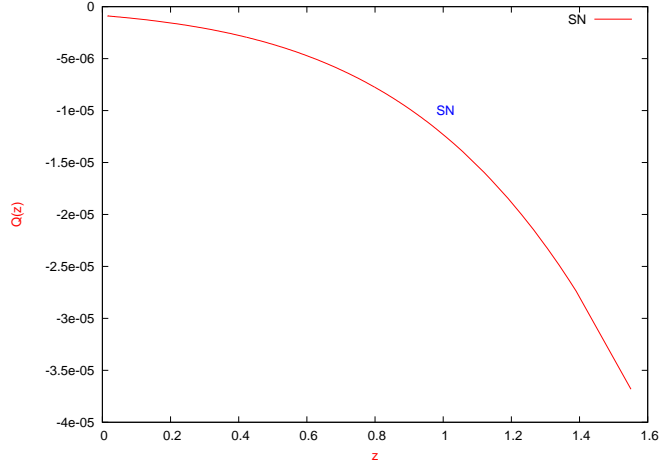}
  \caption{Reconstruction for the interaction function $Q(z)$ as a function of
  the redshift for a spatially flat universe with dark energy interacting with dark
  matter. The reconstruction is derived from the best estimation obtained
  using the type Ia Supernova SCP Union data set sample. Note that the strength of
  the interaction is decreasing to the present.}
  \label{ExchangeSN}
\end{figure}
\end{center}

        \paragraph{\textbf{Cosmological test using Ia Supernova data}}

In what follows we assume the following priors $H_0 = 72 \,\,
\textrm{km} \,\, \textrm{s}^{-1} \, \textrm{Mpc}^{-1}$,
\,$\Omega_{DM}^0 = 0.233$, \,$\Omega_{r}^{0} = 4.63 \times 10^{-5}$,
\, $\Omega_{b}^{0} = 4.62 \times 10^{-2}$. This is motivated by the
fact that current data are converging around these values
\cite{Hinshaw2009}-\cite{Rozo2009}. The density parameter for dark
energy at the present is fixed using the first Friedmann equation
evaluated today $\Omega_{DE}^0 = 1 - \Omega_{DM}^0 - \Omega_{r}^{0}
- \Omega_{b}^{0}$. We constrain the possible values of the remaining
parameters ($w, \hat{\lambda}_{0}, \hat{\lambda}_{1},
\hat{\lambda}_{2}$) using the type Ia Supernova SCP Union data set
(307 SNe-Ia) \cite{Kowalski2008}.

For the SNe Ia test it is defined the observational luminosity
distance \cite{Perlmutter1999,Riess2004} in a flat cosmology as $d_L
(z, w, \hat{\lambda}_{0}, \hat{\lambda}_{1}, \hat{\lambda}_{2}) =
c(1+z)H^{-1}_0 \int_0^z \tilde{H}(z')^{-1} \; dz'$, where
$\tilde{H}(z) \equiv H(z)/H_0$ and $c$ the speed of light. The
\emph{theoretical distance moduli} for the $i$-th supernova with
redshift $z_i$ is $ \mu(z_i)=5\log_{10} [d_L(z_i)/ {\rm Mpc}] +25
\;$. The  statistical function  $\chi^2_{{\rm SNe}}$ becomes
$\chi^2_{{\rm SNe}} (w, \hat{\lambda}_{0}, \hat{\lambda}_{1},
\hat{\lambda}_{2}) \equiv \sum_{k = 1}^{182}
   \left[\mu (z_k) - \mu_k \right]^2 / \sigma_k^2$,
where $\mu_k$ is the \emph{observed} distance moduli for the $k$-th
supernovae and $\sigma_k^2$ is the variance of the corresponding
measurement.

      \paragraph{\textbf{Reconstruction of the parameters}}
We compute the \textit{best estimated values} of the parameters ($w,
\hat{\lambda}_{0}, \hat{\lambda}_{1}, \hat{\lambda}_{2}$) to the
data through $\chi^2$-minimization, using the SNe Ia test. In this
case $z_{max} = 1.551$ corresponding to the farthest Supernova in
the sample used. \\
We obtain as the best estimation: $w = -1.2755$, $\hat{\lambda}_{0}
= -8.5255 \times 10^{-7}$, $\hat{\lambda}_{1} = 7.5755 \times
10^{-9}$ and $\hat{\lambda}_{2} = 5.2755 \times 10^{-10}$, with a
$\chi^2_{\rm{min}}=314.811 \; (\chi^2_{\rm{d.o.f.}}= 1.038)$. For
the age of the universe we have $13.83$ Gyr. The results of the
reconstruction are illustrated in figures \ref{Mixed_SN} and
\ref{ExchangeSN}. From these figures we conclude that:
\begin{itemize}
  \item The density parameters $\Omega_{DM}(z), \, \Omega_{DE}(z)$
  as a function of the redshift are shown in the Fig.
  \ref{Mixed_SN}. We can see that the problem of the cosmic
  coincidence is not solved and in fact it is worse in comparison
  with the $\Lambda$CDM model due to that $Q(z)$ is negative \cite{Campo2009}
  as it is shown in the Fig. \ref{ExchangeSN}.
  We note that if we extrapolate this curve to early times the DE
  density parameter becomes negative which is a serious drawback
  for the existence of an interaction between the dark components.

  \item Fig. \ref{ExchangeSN} shows that the reconstructed interaction
  function $Q(z)$ is always negative signifying an energy transfer
  from DM to DE which worsen the problem of the cosmic coincidence
  in comparison with the $\Lambda$CDM model \cite{Campo2009}.

  \item The preliminary reconstruction shows that, the principal motivation
  for introducing a recent possible interaction between DE and DM
  (to solve the problem of the cosmic coincidence)is not
  supported by recent type Ia Supernova data.
\end{itemize}

Finally we mention that a extended analysis will be presented
elsewhere \cite{Cueva-Nucamendi} which includes a reconstruction of
the interaction term adding recent Cosmic Microwave Background (CMB)
and the Baryon Acoustic oscillations (BAO) data.

        \paragraph{\textbf{Acknowledgments}}
%\begin{theacknowledgments}
     This work is partly supported by grants
     CIC-UMSNH 4.8, PROMEP UMSNH-CA-22, SNI-20733. F.C. thanks to the CONACYT for finantial
     support. We thank to the Numerical Relativity Group of the IFM-UMSNH for the use of its
     computer Cluster for the realization of this work.
%\end{theacknowledgments}

\end{document}